\documentclass[aps,twocolumn,prl,showpacs,color,psfig,epsf]{revtex4}
\usepackage{graphicx}
\usepackage{amsmath,amssymb,latexsym,color,amsfonts}
\bibstyle{apsrev.bib}

\newcommand{\beq}{\begin{equation}}
\newcommand{\eeq}{\end{equation}}

\begin{document}
\title{Shearing active gels close to the isotropic-nematic transition}
\author{M. E. Cates,$^1$, S. M. Fielding,$^2$, D. Marenduzzo$^1$, E. Orlandini$^3$,
J. M. Yeomans$^4$}
\affiliation{$^1$ SUPA, School of Physics, University of Edinburgh,
Edinburgh EH9 3JZ, UK \\
$^2$Manchester Centre for Nonlinear Dynamics and School of Mathematics, University of Manchester, Manchester M13 9EP, UK \\
$^3$ Dipartimento di Fisica and Sezione INFN, Universita' di Padova, Via
Marzolo 8, 31121 Padova, Italy \\
$^4$ The Rudolf Peierls Centre for Theoretical Physics, University
of Oxford, Oxford OX1 3NP, UK}

\begin{abstract}
We study numerically the rheological properties of a slab of active gel close to the isotropic-nematic transition. The flow behavior shows strong dependence on sample size, boundary conditions, and on the bulk constitutive curve, which, on entering the nematic phase, acquires an activity-induced discontinuity at the origin. The precursor of this within the metastable isotropic phase for contractile systems ({\em e.g.,} actomyosin gels) gives a viscosity divergence; its counterpart  for extensile ({\em e.g.,} {\em B. subtilis}) suspensions admits instead a shear-banded flow with zero apparent viscosity.
\end{abstract}

\maketitle

Active gels, including cell extracts and bacterial suspensions, are important and fascinating complex
fluids \cite{ramaswamy,joanny,liverpool,llopis,goldstein,bray,activeLB}. These systems contain self-propelled subunits which, on symmetry grounds,  have generic ordering tendencies that favor an isotropic-nematic (I-N) transition. Unlike conventional nematogens, they are driven by a 
continuous energy flux and remain out of thermodynamic equilibrium even in the steady state.  Active elements contribute to the 
stress as force dipoles that can be characterised as `extensile' or `contractile'. Rodlike extensile particles induce a dipolar flow away from the ends of the rods round to 
their long sides, whereas for contractile rods the flow direction is reversed. This means that an extensile, flow-aligning 
nematogen placed in a shear flow produces a flow that enhances the applied shear; its contractile counterpart opposes the applied shearing. 

In both cases, the quiescent bulk ordered phase (N) is unstable, but restabilizes at high shear rates \cite{ramaswamy}. Confining walls can also restore stability \cite{joanny,activeLB}, creating a finite threshold of activity for onset of spontaneous flow. A recent analysis \cite{liverpool} predicts moreover that the zero-shear viscosity of a contractile
solution should {\em diverge} at the I-N transition. Thereafter, within the N phase, one expects a positive yield stress whose counterpart for an extensile nematic is, however, formally negative \cite{ramaswamy}.
 
In this Letter, we study numerically the steady-state behavior of sheared active gels near the transition to a nematic phase. We discover a shear-banding instability in the extensile case which replaces a negative-viscosity zone by a window of `superfluidity', accommodating a range of macroscopic shear rates at zero stress. 
We also find for contractile systems a viscosity divergence \cite{liverpool}, at the (spinodal) threshold of the I-N transition. Near this, 
we find discontinuous shear thickening at a shear-induced I-N transition: though present in passive nematogens \cite{banding}, the latter 
is strongly perturbed by activity. 
The slab's dimensions and boundary conditions can strongly influence this strikingly rich range of phenomena.

We hope that, as well as offering insights into the highly nonlinear physics of active gels, our numerics will prompt new rheological experiments on soft active 
matter such as bacterial suspensions and actomyosin gels. 
 
To gain our results we used a hybrid lattice Boltzmann (HLB) algorithm \cite{activeLB}, whose approximations are shared by the hydrodynamic equations of motion (detailed below) of active fluids \cite{ramaswamy,joanny,liverpool}; and also a finite difference (FD) algorithm that additionally neglects inertia \cite{banding,fieldingalgo}. 
We assume translational invariance along flow and vorticity directions, reducing the 3D problem to 1D \cite{2D}.

{\em Governing Equations:} In a coarse grained approach the local properties of an (apolar) active fluid can be described by a tensor order 
parameter, $Q_{\alpha\beta}$, whose largest eigenvalue, 2$q$/3, and its associated eigenvector give the magnitude 
and direction of the local orientational order. The equilibrium physics of the passive system is then reproduced by a  
Landau-de Gennes free energy density ${\cal F}$:
\begin{eqnarray}
(1 - \frac {\varphi} {3}) \frac{Q_{\alpha \beta}^2}{2} -
          \frac {\varphi}{3} Q_{\alpha \beta}Q_{\beta
          \gamma}Q_{\gamma \alpha}
+ \frac {\varphi}{4} (Q_{\alpha \beta}^2)^2 +\frac{K}{2}
\left(\partial_{\gamma} Q_{\alpha \beta}\right)^2
\label{eqBulkFree}
\end{eqnarray}
where $\varphi$ controls the magnitude of the ordering. There is a first order transition between the N phase, stable for $\varphi>2.7$ and
the I phase, stable for $\varphi<2.7$. At $\varphi= 3.0$ lies the spinodal point, the limit of the 
metastability of the I phase. In the last term of (\ref{eqBulkFree}) 
$K$ is an elastic constant (in the one constant approximation) \cite{degennes}. 
The equation of motion for $Q_{\alpha \beta}$ is \cite{beris} 
$
D_t Q_{\alpha \beta} = \Gamma H_{\alpha \beta}
$
with $D_t$ a material derivative describing advection by the fluid  velocity $u_\alpha$, and rotation/stretch by flow gradients (see \cite{beris,activeLB}). The molecular field is $H_{\alpha \beta}= -{\delta 
{\cal F} \over \delta Q_{\alpha \beta}} + (\delta_{\alpha \beta}/3) {\mbox {\rm Tr}}{\delta {\cal F} \over \delta Q_{\alpha \beta}}$ 
and $\Gamma$ is a collective rotational diffusivity. 
The fluid velocity obeys the continuity equation, and the Navier-Stokes equation
\begin{eqnarray}\label{navierstokes}
\rho(\partial_t+ u_\beta \partial_\beta) u_\alpha & = &
\partial_\beta (\Pi_{\alpha\beta})+ \eta
\partial_\beta \left (\partial_\alpha u_\beta + \partial_\beta
u_\alpha \right)
\end{eqnarray}
with $\rho$ the fluid density, $\eta$ a Newtonian viscosity, 
and \cite{activeLB}
\begin{eqnarray}
\Pi_{\alpha\beta}= &-&P_0 \delta_{\alpha \beta} +2\xi
(Q_{\alpha\beta}+{1\over 3}\delta_{\alpha\beta})Q_{\gamma\epsilon}
H_{\gamma\epsilon}\nonumber
\end{eqnarray}
\begin{eqnarray}
\label{stress_tensor}
&-&\xi H_{\alpha\gamma}(Q_{\gamma\beta}+{1\over
  3}\delta_{\gamma\beta})-\xi (Q_{\alpha\gamma}+{1\over
  3}\delta_{\alpha\gamma})H_{\gamma\beta}\\ \nonumber
&-&\partial_\alpha Q_{\gamma\nu} {\delta
{\cal F}\over \delta\partial_\beta Q_{\gamma\nu}}
+Q_{\alpha \gamma} H_{\gamma \beta} -H_{\alpha
 \gamma}Q_{\gamma \beta}-\zeta Q_{\alpha\beta}
\end{eqnarray}
The last term of (\ref{stress_tensor}) stems from activity, with $\zeta>0$ (extensile) or  $\zeta<0$ (contractile) \cite{ramaswamy,joanny,liverpool,activeLB}.
The parameter $\xi$ depends on the geometry of active elements; it 
determines whether the material is flow-aligning or flow-tumbling.  We restrict 
ourselves to the former by taking $\xi = 0.7$.
Our equations describe for simplicity apolar active gels, whereas many  such gels ({\em e.g.}, actomysin) are polar; however, we expect 
similar rheology, as the hydrodynamic active stress has the same form in both cases \cite{ramaswamy}. Rheological experiments might also be possible on apolar active matter, 
such as melanocytes or vibrated rods
\cite{ramaswamy}.

\begin{figure}
\centerline 
{\includegraphics[width=3.2in]{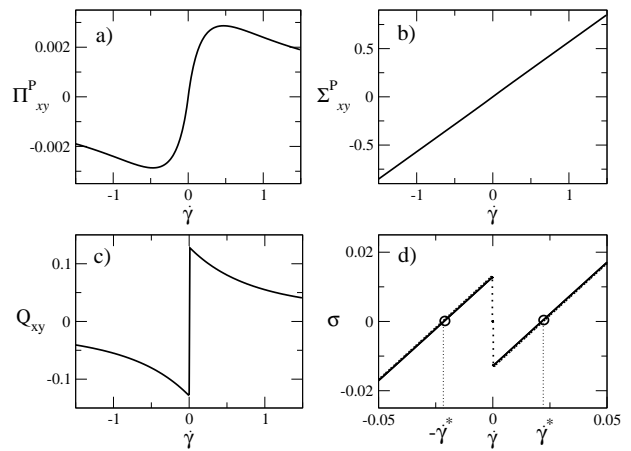}}
\caption{For an extensile nematic ($\varphi = 3.0,\xi = 0.7,\eta = 0.567, \zeta = 0.1$), as functions of $\dot\gamma$ in steady state :(a) Passive order-parameter stress $\Pi^P_{xy} =\Pi_{xy}+\zeta Q_{xy}$; (b) Total passive stress $\Sigma^P_{xy}= \Pi^P_{xy}+\eta\dot\gamma$; (c) Order parameter $Q_{xy}$; (d) Total shear stress $\sigma(\dot\gamma)$ (dotted curve: the same for $\varphi = 2.95$.)} 
\label{f1}
\end{figure}

{\em Constitutive Curves:} We first compute \cite{flowcurveshow} curves $\sigma(\dot\gamma)$ for the shear stress $\sigma$ at {\em uniform} imposed shear rate $\dot\gamma \equiv \partial_y u_x$. Fig.\ref{f1} shows this curve (and its stress contributions) for an extensile nematic. With our parameters, the passive terms in $\Pi_{xy}$  are much smaller than the viscous term $\eta\dot\gamma$,  so $\sigma \simeq \eta\dot\gamma - \zeta Q_{xy}(\dot\gamma)$. In both contractile and extensile nematics $Q_{xy}$ has an upward discontinuity at the origin (see Fig.\ref{f1}); for contractile materials ($\zeta <0$) this creates an upward step in $\sigma(\dot\gamma)$, or
yield stress \cite{ramaswamy}. The constitutive curve for the extensile case ($\zeta > 0$, Fig.\ref{f1}) instead has a downward step, or `negative yield stress', causing a zone of negative viscosity ($\sigma/\dot\gamma<0$) for $|\dot\gamma|<\dot\gamma^*$, which can be circumvented by splitting into shear bands of $\dot\gamma = \pm \dot\gamma^*$. 
\if{
(Only this choice avoids negative viscosity in either band). 
}\fi
The resulting composite flow curve $\sigma(\dot\gamma)$, with $\dot\gamma$ now the applied (mean) shear rate, connects $\dot\gamma = \pm \dot\gamma^*$ with a horizontal `tie line' at zero stress. 
The stability of such bands is far from certain, especially if translational invariance is not assumed. However, our FD work (data not shown) confirms the existence of such states in the 1D bulk (large $L$) limit. Figs.~5-7 of \cite{activeLB} on confined slabs also offer evidence for this N/N banding scenario which, we argue below, {\em should also control flow of extensile systems initialized in the I phase} close the I-N spinodal. 

{\em Slab Geometry:} We next consider a slab of active gel between parallel plates at $y$-separation $L$, with $x$-velocities $\pm \dot\gamma L/2$.   Most of our HLB simulations were performed with $L=100$, $\Gamma=0.33775$, $K=0.04$, and $\zeta= \pm
0.005$. These parameter values, and our results, are reported in LB units; such units map {\em e.g.} to $\Delta x=0.05$ $\mu$m and $\Delta t=0.67$ $\mu$s with $\Gamma = 1$ Poise, $\eta = 1.1$ Poise, $K = 10 pN$ \cite{degennes} and $\zeta = 500$ Pa. The only length scale in Eqs.(\ref{eqBulkFree}-\ref{stress_tensor}) is then $\ell \equiv K^{1/2} \simeq 5$nm. (This sets the scale for band interfaces, although these can be broader close to any second-order transitions.)
Our HLB parameter represent molecular nematogens in a microfluidic-device-scale slab a few $\mu$m across \cite{activeLB} although very different values might arise in some biological contexts. 

Below we compare `fixed' (anchored) boundary conditions, in which $Q_{\alpha\beta}$ is specified on the walls,  and `free' ones which effectively set $\partial_yQ_{\alpha\beta}=0$, at $y=0,L$. 
For shear-banding problems, the latter choice promotes convergence to bulk behavior \cite{wallstuff}; our FD work uses this, and to the same end, chooses small $\ell/L =5\times 10^{-4}$. The boundary condition on $u_\alpha$ is taken as no-slip.

{\em Linear Contractile Rheology:} We now report apparent viscosities  $\eta_{\rm app}(\dot\gamma,L,...) = (\Pi_{xy}+\eta \partial_y u_x)/\dot\gamma$. For homogeneous flows, these give the bulk flow curve $\eta_{\rm app} =\sigma(\dot\gamma)/\dot\gamma$ and reduce to the zero-shear viscosity as $\dot\gamma\to 0$. 

\begin{figure}
\centerline {\includegraphics[width=3.7in]
{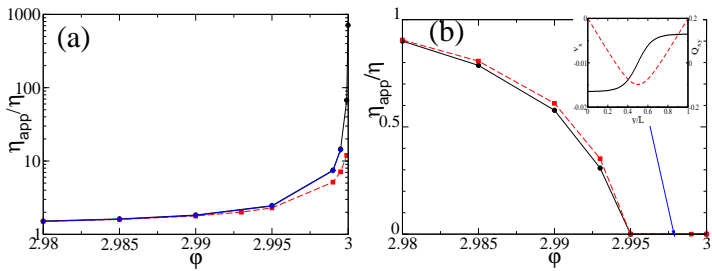}}
\caption{(a) Apparent zero shear viscosity for an isotropic 
contractile gel in a slab under  shear (solid and dashed lines). Solid curve(s): free boundary conditions; HLB data are indistinguishable from those derived from the bulk flow curve \cite{flowcurveshow}. Dashed curve: fixed boundaries (director along flow).  (b) 
Apparent viscosity in a linear regime for an active 
extensile gel in a slab under  shear. Solid and dashed curves
correspond to systems with free and fixed boundaries respectively.
The insets show plots of $Q_{xy}$ and $u_x$ (solid and dashed lines) in
the $\eta_{\rm app}=0$ phase.}
\label{linear_rheology}
\end{figure}

We start with the zero-shear viscosity of a contractile gel in the metastable I phase, $2.7 < \varphi \le 3.0$. We observe a viscosity divergence at the latter (spinodal) point, as predicted in \cite{liverpool} which, however, addressed strictly 2D nematogens for which binodal and spinodal coincide (as there is then no cubic term in (\ref{eqBulkFree})). 
This is shown in Fig.~\ref{linear_rheology}a, where the divergence is seen with free boundary conditions, for which the flow remains homogeneous, but is suppressed if $Q_{\alpha\beta}$ is anchored at the walls with director along the flow (creating nonuniform shear near the walls).  In the latter case, $\eta_{\rm app}$ increases linearly with $L$ (data not shown; we have simulated up to $L=400$) until it saturates, for $\varphi<3$, at the free-boundary value. To understand this, note that surface anchoring lifts the rotational degeneracy of the N phase, so that deviations of the director $\bf n$ away from $x$ feel a finite restoring force (that vanishes only as $L\to\infty$). Hence the active shear stress, $\sim -\zeta q n_xn_y$, remains of order $\dot\gamma$ and there can be no yield stress, nor any viscosity divergence.

{\em Extensile Rheology:} Consider next systems initialized in the metastable I phase of an {\em extensile} material ($\zeta = 0.005$). These have strikingly different behavior, shown in 
Fig.~\ref{linear_rheology}b. Not only does $\eta_{\rm app}$ decrease strongly as the
spinodal is approached: it becomes strictly zero in an interval $\varphi_c(\zeta,L)<\varphi<3$. We argue that this `superfluid' window replaces one of negative viscosity on the constitutive curve, arising for $\varphi > \varphi^*(\zeta)$ (say), and similar to that already discussed for extensile nematic phases (Fig.~\ref{f1}).  To understand its continued presence here, note that as $\varphi\to 3^-$ and at $\dot\gamma = 0$, $\partial Q_{xy}/\partial\dot\gamma \to -\infty$; hence at small $|\dot\gamma|$ the flow curve has large negative slope (precursor to the step discontinuity in the nematic). As $|\dot\gamma|$ increases, nonmonotonic features can arise, but the flow curve then joins that of the nematic, {\em before} stability is restored at $|\dot\gamma| = \dot\gamma^*$. Close enough to the spinodal, therefore, the flow curve of the metastable I phase still has all the features needed for formation of bulk N/N shear bands. We then expect $\varphi_c(\zeta,L\to\infty) = \varphi^*(\zeta)$; this is numerically consistent with our HLB and FD data. 

However, bulk shear bands do not provide a complete picture for the confined slab of Fig.~\ref{linear_rheology}b, where $Q_{xy}$ and $u_x$ remain relatively smooth on the scale of $L$ (inset). With fixed boundary conditions, the zero-viscosity window is narrower and can disappear at small $L$. A complementary interpretation of our N/N-banded state is as a `spontaneous flow' phase, predicted inside the nematic region  of confined extensile materials \cite{joanny,activeLB}. Our results suggest that, on increasing $\varphi$, a direct transition from metastable isotropic to the spontaneous flow phase is possible.

\begin{figure}
\centerline {\includegraphics[width=3.in]
{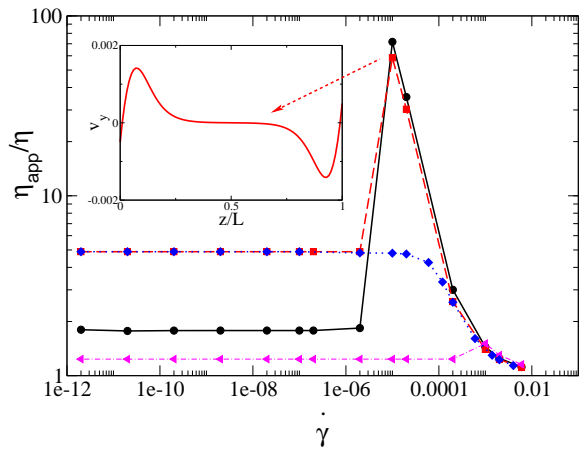}}
\caption{Apparent viscosity vs. $\dot\gamma$ for
$\varphi = 2.99$ (solid, dashed and dotted line) and $\varphi = 2.8$ (dot-dashed line) for contractile gels under shear. Anchored BCs are considered, with $Q_{\alpha\beta} = 0$ (solid, dot-dashed) or with ${\bf n} \parallel \hat{\bf x}$ (long-dashed and dotted lines). 
For the dotted line, the bulk was also initialised with nematic order.} 
\label{nonlinear_contractile}
\end{figure}

{\em Nonlinear Contractile Rheology:}
We again focus on the metastable I regime, $2.7 < \varphi \le 3.0$,  whose passive counterpart shows shear thinning, and a shear-induced I-N
transition \cite{beris}. 
%
In Fig. \ref{nonlinear_contractile} we plot HLB results for $\eta_{\rm app}(\dot\gamma)$ in a confined slab of a contractile gel, with each shear rate $\dot\gamma$ initiated from rest. We find a rich phenomenology, with strong dependences on
boundary and initial conditions. Consider first the solid line, for $\varphi=2.99$, with an initially
isotropic state, and fixed $Q_{\alpha\beta} = 0$ on the walls. For small $\dot\gamma$ there is a linear regime of constant
viscosity; the material remains in the I phase.  There is then a sharp increase in $\eta_{\rm app}$, {\em i.e.,} strong shear thickening (discontinuous at large $L/\ell$), with onset of nematic order oriented near but above the Leslie angle $\theta = \theta^*$, and a drastically non-Newtonian flow profile (Fig. \ref{nonlinear_contractile}, inset). At larger $\dot{\gamma}$, $\theta$ decreases and $\eta_{\rm app}$ falls towards $\eta$. 
The dashed curve shows the same system but with fixed nematic ordering and ${\bf n} \parallel \hat{\bf x}$ at the walls. For small $\dot\gamma$ the shear flow is now mainly confined to wall layers; the director has $\theta \simeq 0$ throughout the slab. In the wall layers, the backflow term $\Pi_{\alpha\beta}$ acts to restore the ordering, creating extra dissipation and enhanced $\eta_{\rm app}$ at small $\dot\gamma$. 
(This scenario is reminiscent of a passive nematogen flowing in the permeation mode close to the isotropic-cholesteric transition \cite{permeation}, where
energy from the applied shear is converted into a flow that opposes the shear and reduces the order.)
Shear thickening is now accompanied by an abrupt director transition,  
near the slab centre. As $\dot\gamma$ is raised, $\theta$ decreases, $\eta_{\rm app}$ falls, and a Newtonian flow profile is restored. If, instead, the system is initialised in the nematic phase (dotted curve) the jump in $\theta$ disappears, as does the peak in $\eta_{\rm app}$. Finally, we consider an initial I state further from the spinodal at $\varphi = 2.8$, with $Q_{\alpha\beta}=0$ at the wall (dot-dashed line). This system remains isotropic for longer: $\eta_{\rm app}$ is smaller in the
linear regime, remains there to higher $\dot{\gamma}$, and shows a smaller peak at the shear-induced I-N transition.

\begin{figure}
\centerline {\includegraphics[width=2.8in]
{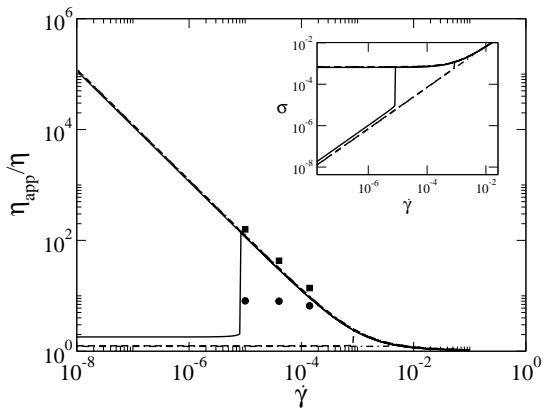}}
\caption{Viscosity and (inset) flow curves for active contractile gels in bulk homogeneous shear: $\varphi = 2.8, 2.9, 2.99$ (dot-dashed, dashed, solid). Vertical line segments signify the end of the I branch and a jump up to N. Data 
with nematic anchoring as in Fig. \ref{nonlinear_contractile}
is shown for an initially isotropic (squares) or nematic (circles) bulk.\\}
\label{f2}
\end{figure}

These results for confined slabs are again illuminated by computing  the homogeneous bulk flow curve $\sigma(\dot\gamma)$ \cite{flowcurveshow}. For $\zeta < 0$ this has a low viscosity I branch, terminating at some finite flow rate, and a high viscosity, shear thinning N branch (reflecting the presence of a yield stress). Fig.~\ref{f2} shows both the flow curves (inset) and the viscosity $\eta(\dot\gamma)$. Discontinuous shear-thickening arises on tracking up to the end of the I branch, then jumping vertically to N. (On decreasing shear rate, N can be tracked downwards to low $\dot\gamma$ and likely remains metastable.) Once again, there are features of the confined behavior, such as boundary-condition sensitivity, that this bulk picture cannot explain. 

\begin{figure}
\centerline {\includegraphics[width=2.5in]{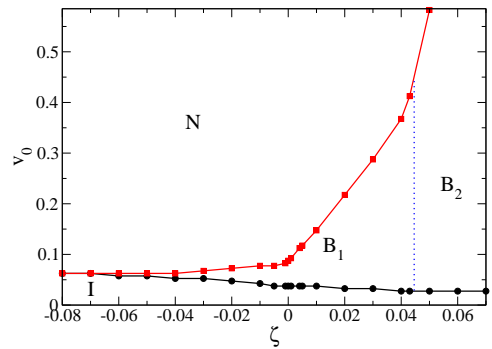}}
\caption{Dynamic phase diagram in the $(\zeta,v_0)$ plane (with $v_0 = \dot\gamma L$ the wall velocity; $L = 100$) for an active liquid
crystal close to the isotropic-nematic transition ($\varphi=2.695,\eta = 0.28$).
The phases are labelled as follows: 
N= nematic, I = isotropic, B1 = I/N banded, B2 = N/N banded.}
\label{shear_banding}
\end{figure}

We finally consider the effect of activity on the I/N shear banding present already in passive nematogens \cite{banding}. To this end, we fix $\varphi=2.695$ and
$K=0.01$; in the passive system ($\zeta=0$) there is a range of shear rates $\dot\gamma$ for which the fluid splits into I and N bands of high and low viscosity.  Activity has a significant impact: as shown in Fig. \ref{shear_banding}, it widens or narrows
the banding window according to the sign of $\zeta$. (For larger $\phi$ the bands pinch off giving a direct I/N transition, as reported above, in sufficiently contractile materials.) The trend concurs with expectations from the bulk flow curve: $\sigma(\dot\gamma)$ for the N phase is lowered by extensile and raised by contractile activity, allowing the former system to band more easily.\\

To conclude, we have solved the hydrodynamic equations of motion of an active nematogen close to the I-N transition subjected to a shear flow, assuming translational invariance along flow and vorticity directions. For contractile gels and  free boundary conditions there is a divergence of the apparent viscosity at the I-N spinodal \cite{liverpool}. Extensile gels, in contrast, enter a zero-viscosity phase of N/N shear bands as the spinodal is approached. In the nonlinear regime, both extensile and contractile materials show non-monotonic effective flow curves, with details strongly dependent on initial and boundary conditions. Finally, extensile (contractile) activity 
stabilizes (suppresses) I/N shear banding. We hope our results will spur experimental investigations of the rheology of active gels, such as
extensile bacterial suspensions, or semidilute contractile actomyosin or kinesin-microtubule solutions. \\

SMF thanks EPSRC EP/E5336X/1 for funding; MEC holds a Royal 
Society Research Professorship.


\begin{thebibliography}{99}
\bibitem{2D} 2D HLB studies will be reported elsewhere.
\bibitem{flowcurveshow} $\sigma(\dot\gamma)$ for putative homogeneous bulk flow was found by Euler time integration with no spatial coordinates.
\end{thebibliography}
\end{document}